\documentclass[12pt]{article}

\usepackage[left=1.25in,top=1.0in,right=1.25in,bottom=1.0in]{geometry}
\usepackage{amsfonts}
\usepackage{amssymb}
\usepackage{graphicx}
\usepackage{setspace}
\usepackage{amsmath}
\usepackage{amssymb}
\usepackage{amsbsy}
\usepackage{amsthm}
\usepackage{paralist}
\usepackage{bm}

\usepackage{mathtools}
\usepackage{cases}
\usepackage{algorithm}
\usepackage{algpseudocode}

\newtheorem{mydefn}{Definition}

\def\E{\qopname\relax o{E}}

\newcommand{\bbeta}{{\beta}}

\newcommand{\N}{\mbox{N}}
\newcommand{\PG}{\mbox{PG}}

\newcommand{\Ga}{\mbox{Ga}}

\title{Expectation-maximization for logistic regression}

\author{
James G.~Scott\footnote{McCombs School of Business
and Division of Statistics and Scientific Computing;
james.scott@mccombs.utexas.edu} \\
Liang Sun\footnote{ Department of Operations Research and Industrial Engineering; 
sallylia@gmail.com}\\
\\
University of Texas at Austin
}

\date{First version: September 2012\\
This version: May 2013}

\begin{document}

\maketitle

\begin{abstract}
We present a family of expectation-maximization (EM) algorithms for binary and negative-binomial logistic regression, drawing a sharp connection with the variational-Bayes algorithm of \cite{jaakkola:jordan:2000}.  Indeed, our results allow a version of this variational-Bayes approach to be re-interpreted as a true EM algorithm.  We study several interesting features of the algorithm, and of this previously unrecognized connection with variational Bayes.  We also generalize the approach to sparsity-promoting priors, and to an online method whose convergence properties are easily established.  This latter method compares favorably with stochastic-gradient descent in situations with marked collinearity.  This paper summarizes our methods and conclusions, with details of experiments provided in a supplemental file.

\end{abstract}

\section{Introduction}

Consider a logistic regression, where $y_t \sim \mbox{Binom}(m_t, w_t)$ for a fixed number $m_t$ of binary trials, $t = 1, \ldots, N$; and where the log-odds of success are modeled as a linear function of $d$ predictors:
$$
\psi_t = \log \left( \frac{w_t}{1-w_t} \right) = x_t^T \beta \, .
$$
Suppose further that $\beta$ is given a normal prior with mean $\mu$ and covariance $\Sigma$, and that our goal is to estimate, or approximate, the posterior distribution of $\beta$.

The purpose of this article is to derive the solution to this problem embodied by Algorithm \ref{alg:ig1}, and to study both its subsequent elaborations and its connection with variational-Bayes inference.  The algorithm iterates between two simple steps: (1) use the current value of $\beta$ to update a diagonal matrix $\Omega$ of working parameters $\{\omega_t\}$, and (2) use the new $\Omega$ to construct and solve a set of normal equations for the new $\beta$.  Using recently developed distributional theory from \cite{polson:scott:windle:2012a}, it is straightforward to show that these two steps form an exact expectation-maximization (EM) algorithm.  The $\omega_t$'s play the role of notionally missing data; their conditional distribution, which falls within the family of infinite convolutions of gammas, will be described shortly.  The ascent property of the EM algorithm, together with the log-concavity of the posterior distribution, guarantee that the sequence of iterates $\{\beta^{(1)}, \beta^{(2)}, \ldots\}$ converges to the posterior mode.

\begin{algorithm}[t]
\begin{algorithmic}
\State Data: pairs $(m_{t}, y_t)$ for $t = 1, \ldots, N$; design matrix $X$ having $x_t^T$ as row $t$.
\State Prior: $\beta \sim \N(\mu, \Sigma)$.
\State Define: \texttt{solve($S, d$)}, a routine which solves the system $S \beta = d$
\State $\kappa \leftarrow (y_1 - m_1/2, \ldots, y_N - m_N/2)^T$; $d \leftarrow X^T \kappa $
\Repeat 
\State \textsc{For} $t = 1, \ldots, N$:  $\psi_t \leftarrow x_t^T \beta$; $\omega_t \leftarrow \frac{m_t}{2 \psi_t} \tanh(\psi_t/2)$
\State $\Omega \leftarrow \mbox{diag}(\omega_1, \ldots, \omega_N)$;  $S \leftarrow X^T \Omega X $
\State $\beta \leftarrow \texttt{solve}(S + \Sigma^{-1}, d +  \Sigma^{-1} \mu)$
\Until $\beta$ converges.
\end{algorithmic}
\caption{Expectation-maximization for logistic regression.}
\label{alg:ig1}
\end{algorithm}

The details of Algorithm \ref{alg:ig1} may be familiar to many readers in a different guise, particularly the functional form of the update for the ``missing'' $\omega_t$.  This is not an accident: our method is closely related to the variational-Bayes approach to logistic regression described by \cite{jaakkola:jordan:2000}.  Indeed, one way of arriving at Algorithm \ref{alg:ig1} is via a pure variational argument appealing to convex duality.  But we pursue an entirely different, probabilistic line of argument, giving rise to subtle and importance differences from the typical variational-Bayes approach.

Section 2 introduces the method.  Section 3 pursues the connection with variational Bayes.  Section 4 describes online and sparse variants of the algorithm.  A supplemental file contains a large suite of numerical experiments, beyond the few described in the main paper.

\section{Construction of the EM algorithm}

\subsection{The Polya-Gamma family and logistic likelihoods}

Our EM algorithm exploits a latent-variable representation of logistic likelihoods introduced by \cite{polson:scott:windle:2012a} and further explored in \cite{pillow:scott:2012} and \cite{zhou:etal:2012}.  This includes both the logit and negative-binomial models as special cases.  These are likelihoods that involve products of the form
\begin{equation}
 \label{eqn:logitlikelihood-general}
L_t = \frac{(e^{\psi_t})^{a_t}}{(1+e^{\psi_t})^{b_t}} \, ,
\end{equation}
where $\psi_t$ is a linear function of parameters, and where $a_t$ and $b_t$ involve the response for subject $t$.
The authors of \cite{polson:scott:windle:2012a} exploit these facts to derive an efficient Gibbs sampler for the logit model.  In contrast, our focus is on using the same representation to derive expectation-maximization algorithms.

The key result is that terms of form (\ref{eqn:logitlikelihood-general}) are mixtures with respect to a Polya-Gamma distribution.  Define $\omega \sim PG(b,0)$, $b > 0$ as the infinite
convolution of gammas having Laplace transform
\begin{equation}
\label{eqn:PGmgf1}
\E\{\exp(- \omega t)\} = \prod_{i=1}^t \Big(1 + \frac{t}{2 \pi^2
  (k-1/2)^2}\Big)^{-b} = \frac{1}{\cosh^b(\sqrt{t/2})} \, .
\end{equation}
The second equality arises from the fact that the hyperbolic cosine function is holomorphic over the entire complex plane, and according to the Weierstrass factorization theorem, may therefore be represented as a product in terms of its zeros.  This Laplace transform is easily inverted by recognizing each term in the product as the Laplace transform of a Gamma distribution.  We therefore conclude that if $\omega \sim \PG(b,0)$, then it is equal in distribution to an infinite sum of gammas:
\begin{eqnarray*}
  \omega &\stackrel{D}{=}&  \frac{1}{2 \pi^2} \sum_{k=1}^{\infty} \frac{g_k }{(k-1/2)^2} \, ,
\end{eqnarray*}
where each $g_k$ is an independent Gamma$(b,1)$ random variable.  The general $\PG(b,c)$ class is constructed via exponential tilting of the $\PG(b,0)$ density:
\begin{equation}
\label{eqn:polyagamma.generalpdf}
p(\omega \mid b, c) \propto \exp \left( -\frac{c^2}{2} \omega \right) p(\omega \mid b,0) \, ,
\end{equation}
The corresponding Laplace transform may be calculated and inverted by a similar path.  We omit the details of this calculation, which leads directly to the following definition.
\begin{mydefn}
A random variable $X$ has a Polya-Gamma distribution with parameters $b > 0$ and $c \in \mathcal{R}$, denoted $X \sim \PG(b,c)$,  if
\begin{equation}
\label{eqn:PGdistribution1}
X \stackrel{D}{=} \frac{1}{2 \pi^2} \sum_{k=1}^{\infty} \frac{g_k}{(k-1/2)^2 + c^2/(4\pi^2)} \, ,
\end{equation}
where each $g_k \sim \Ga(b,1)$ is an independent gamma random variable.
\end{mydefn}

To make matters concrete, we take the case of binomial logistic regression, where we observe $N$ triplets $D = \{(m_t, y_t, x_t)\}$, respectively denoting the number of trials, the number of successes, and the predictors for case $t$.  In many data sets we simply have $m_t = 1$ and $y_t$ either 0 or 1, but this need not be the case.  Indeed, the negative-binomial case arises when $m_t = y_t + r$ for some overdispersion parameter $r$.  For all models of this form, the likelihood in $\bbeta$ is
$$
L(\bbeta) = \prod_{t=1}^N \frac{(e^{\psi_t})^{y_t}}{(1+e^{\psi_t})^{m_t}} 
$$
where $\psi_t = x_t^T \bbeta$ is the linear predictor.  The fundamental integral identity arising from the Polya-Gamma data-augmentation trick (see \cite{polson:scott:windle:2012a}) allows us to rewrite each term in the likelihood as follows.
\begin{eqnarray}
\frac{ \{\exp(\psi_t) \}^{y_t} } { \{1+ \exp(\psi_t) \}^{m_t} }
  \propto e^{\kappa_t \psi_t} \int_0^{\infty} e^{-\omega_t \psi_t^2 / 2} \  p(\omega \mid m_t, 0) \ d \omega \label{eqn:likemixture} \, ,
\end{eqnarray}
where $\kappa_t = y_t - m_t/2$, and where the mixing distribution is
Polya-Gamma; this identity arises from evaluating the Laplace transform (\ref{eqn:PGmgf1}) at $\psi^2$.   Moreover, the conditional distribution for $\omega_t$ that arises in treating the above integrand as a joint density is simply an exponential tilting of the prior, and is therefore also in the Polya-Gamma family: $(\omega \mid \psi) \sim \PG (b, \psi)$.

Appealing to (\ref{eqn:likemixture}), the complete-data log-likelihood $Q(\bbeta)$, given all $\omega_t$, is a quadratic form in $\bbeta$.  This yields a conditionally Gaussian likelihood:
\begin{equation}
\label{eqn:condgausslike}
Q(\bbeta) = -\frac{1}{2} \sum_{t=1}^N \omega_t (x_t^T \bbeta)^2  +  \sum_{t=1}^N \kappa_t x_t^T \bbeta \, .
\end{equation}
This expression is linear in $\omega_t$, meaning that the EM algorithm has an especially simple structure.  In the E step, given the current estimate $\hat{\bbeta}$, we compute $E\{Q(\bbeta) \mid \hat{\bbeta}\} = Q(\bbeta \mid \hat{\omega}_1, \ldots, \hat{\omega}_{N})$, where $\hat{\omega}_t$ is the conditional expected value of $\omega_t$, given the data and the current iterate for $\bbeta$.  We describe this calculation in detail below. Meanwhile, in the $M$ step, we choose the new $\hat{\bbeta}$ to maximize $E\{Q(\bbeta \mid \hat{\bbeta})\}$.  This expression looks like the kernel of a normal distribution. Collecting terms and completing the square, we find that the new value of $\hat{\bbeta}$ maximizes a quadratic form involving complete-data sufficient statistics $S$ and $d$:
$$
Q(\bbeta \mid \omega_1, \ldots, \omega_{N}) = -\frac{1}{2} \bbeta^T S \bbeta + \beta^T d \quad \mbox{where} \quad S = X^T \Omega X \; \mbox{and} \; d = X^t \kappa \, .
$$
Here $\Omega$ is the diagonal matrix $\mbox{diag}(\omega_t)$, and $\kappa$ is the column vector $(\kappa_1, \ldots, \kappa_n)$.  Thus $\hat{\bbeta}$ solves the linear system $S \hat{\bbeta} = d$.  We have carried through the calculation without accounting for the contribution of the prior.  But the log prior density is also a quadratic form in $\bbeta$ and therefore combines easily with the above to yield the complete-data posterior distribution:
\begin{equation}
\label{eqn:completedataposterior}
p_c(\bbeta) = \N(m_\bbeta, V_\bbeta) \quad \mbox{where} \quad V_\bbeta^{-1} = S + \Sigma^{-1} \quad \mbox{and} \quad  m_\bbeta = V_{\bbeta}(d + \Sigma^{-1} \mu) \, .
\end{equation}

If we wish, we can solve this system directly at each step. Alternatively, if this is too costly, we can simply take a step in the right direction toward the solution starting from the previous value using, for example, the linear conjugate-gradient algorithm.   This will typically be much faster than a full solve, and will still lead to global convergence, as it is sufficient to take a partial M step that merely improves the observed-data objective function.

While the density of a Polya-Gamma random variable can be expressed only as an infinite series, its expected value may be calculated in closed form.  To see this, apply the Weierstrass factorization theorem once more to calculate the Laplace transform of a $\PG(b,c)$ distribution as\begin{align}
\label{eqn:pg-mgf}
 \E_{\omega} \left\{ \exp\left( - \omega t \right) \right\} 
  &=  \frac{\cosh^{b} \left( \frac{c}{2} \right)}  {\cosh^{b} \left(
      \sqrt{\frac{c^2/2+t}{2}} \right) }  \quad = \quad \prod_{k=1}^\infty \left( \frac{1+\frac{c^2/2}{2(k-1/2)^2\pi^2}}
  {1+\frac{c^2/2+t}{2(k-1/2)^2\pi^2}} \right)^{b}  \\
 & = \prod_{k=1}^\infty (1+ d_k^{-1} t )^{-b} \; , \quad  {\rm where} \;  d_k =
 2\left(k-\frac{1}{2}\right)^2\pi^2 + c^2/2 \; . \notag
\end{align}

Differentiating this expression with respect to $t$, negating, and evaluating
the result at $t=0$ yields $\E(\omega) = \frac{b}{2c} \tanh(c/2)$ for a $\PG(b,c)$ random variable $\omega$.  Therefore, the E step in our EM algorithm simply involves computing
\begin{equation}
\label{eqn:omegahat2}
\hat{\omega}_t = \left(\frac{m_t}{2 \hat{\psi}_t}\right) \tanh(\hat{\psi}_t/2) \, \quad \, \mbox{with} \quad \hat{\psi}_t = x_t^T \hat{\bbeta} \, ,
\end{equation}
and using these to construct the complete-data sufficient statistics previously defined.  The fact that the algorithm converges to the posterior mode follows trivially from the ascent property of EM algorithms, together with the log-concavity of the posterior distribution.

%

\subsection{QN-EM: Quasi-Newton acceleration}

Quasi-Newton acceleration is a powerful technique for speeding up an EM algorithm.  Suppose that we decompose the observed-data log posterior as $L(\bbeta) = C(\bbeta) - R(\bbeta)$, where $C(\bbeta)$ is the complete-data contribution arising from the Polya-Gamma latent-variable scheme, and $R(\bbeta)$ is the remainder term.  The corresponding decomposition of the Hessian matrix is: $-\nabla^2 L(\bbeta) = -\nabla^2 C(\bbeta) + \nabla^2 R(\bbeta)$; the fact that $-\nabla^2 R(\bbeta)$ is a non-negative definite matrix follows from the information inequality.  As the complete-data log-posterior is Gaussian, $-\nabla^2 C(\bbeta)$ is the inverse of the covariance matrix given in (\ref{eqn:completedataposterior}).  The idea of quasi-Newton acceleration is to iteratively approximate the Hessian of the remainder term, $\nabla^2 R(\bbeta)$, using a series of inexpensive low-rank updates.  The approximate Hessian $\tilde{H} =  \nabla^2 L(\bbeta) - \nabla^2 R(\bbeta)$ is then used in a Newton-like step to yield the next iterate, along with the next update to $\nabla^2 R(\bbeta)$.  Because both Newton's method and the M step of the EM algorithm require solving an order-$d$ linear system, the only extra per-iteration cost is forming a (comparatively cheap) low-rank update to the approximation for $\nabla^2 R(\bbeta)$.   See \cite{lange:1995} for a full explanation of the general theory of quasi-Newton acceleration.

A few general conclusions emerge from our experiments.  First, iteratively re-weighted least squares often fails to converge, especially when initialized from a poor location.  As many others have observed, this reflects the numerical instability in evaluating the true Hessian matrix far away from the solution.  Third, the EM algorithm is robust due to the guaranteed ascent property, but slow.  Finally, the QN-EM algorithm is equally robust, but far faster.  On our experiments, it usually required between 10 and 100 times fewer iterations than the basic EM to reach convergence.

\section{The connection with variational-Bayes inference}

We now draw a sharp connection with the variational algorithm for Bayesian logistic regression described by \cite{jaakkola:jordan:2000}.  Let $l_t(\bbeta)$ denote the contribution to the log-likelihood of case $t$:
\begin{eqnarray*}
l_t(\bbeta) &=& y_t \psi_t - m_t \log \{ 1 + \exp(\psi_t) \} \\
&=& (y_t - m_t/2) \psi_t - m_t \log \{ \exp(\psi_t/2) + \exp(-\psi_t/2) \} \, ,
\end{eqnarray*}
where $\psi_t = x^T \bbeta$ is implicitly a function of $\bbeta$.  The second term, $\phi(\psi_t) = \log \{ \exp(\psi_t/2) + \exp(-\psi_t/2) \}$, is symmetric in $\psi_t$ and concave when considered as a function of $\psi_t^2$.  Appealing to standard results from convex analysis, we may therefore write it in terms of its Legendre dual $\phi^{\star}$:
$$
\phi(\psi_t) = \inf_{\lambda_t} \left\{ \lambda_t \psi_t^2  - \phi^{\star}(\lambda_t)  \right\} =  \hat{\lambda}(\psi_t) \psi_t^2  - \phi^{\star}\{ \hat{\lambda}(\psi_t)\} \, ,
$$
where $\phi^{\star}(\lambda_t) = \inf_{\xi_t} \{ \lambda \xi_t^2 - \phi(\xi_t) \} $.  Moreover, by differentiating both sides of the above equation, the optimal value for $\lambda_t$ as a function of $\psi$, is easily shown to be
$$
\hat{\lambda}(\psi) = \frac{ \phi'(\psi)} {2 \psi} =  \frac{1}{4\psi} \tanh(\psi/2) \, .
$$

This leads to the following variational lower bound for $l_t(\bbeta)$:
\begin{equation}
\label{eqn:variationalsuffstat}
l_t(\bbeta) \geq f_t(\bbeta, \xi_t) = (y_t - m_t/2) (\psi_t - \xi_t) - m_t \hat{\lambda}(\xi_t)(\psi_t^2 - \xi_t^2) + l_t(\xi_t) \, .
\end{equation}
This holds for any choice of $\xi_t$, with equality achieved at $\xi = \psi_t = x^T \bbeta$.  Upon defining $\omega_t \equiv m_t \lambda_t/2$ and $\kappa_t \equiv y_t - m_t/2$, it is readily apparent that each summand in (\ref{eqn:condgausslike}) is identical to the expression just given in (\ref{eqn:variationalsuffstat}), up to an additive constant involving $\xi$.  In each case, the approximate posterior takes the Gaussian form given in (\ref{eqn:completedataposterior}).

This provides a probabilistic interpretation to the purely variational argument of \cite{jaakkola:jordan:2000}.  It is especially interesting that we are able to identify $\lambda$, the argument of the dual function $\phi^{\star}(\lambda)$, as a rescaling of the missing data $\omega_t$ that arise from a complete different line of argument under the Polya-Gamma EM approach.  Both terms play the role of the inverse-variance in a conditionally Gaussian likelihood.  For further insight, compare the expression for the optimum value of $\lambda$ with the expression for the conditional expected value of $\omega_t$ in Formula (\ref{eqn:omegahat2}) and the inner loop of Algorithm \ref{alg:ig1}.

Despite this equivalence of functional form, the EM and variational-Bayes algorithms do not give the same answers.  The former converges to the posterior mode and treats $\lambda_t$ as missing data having a Polya-Gamma mixing distribution, while the latter operates by a fundamentally different inferential principle.  From Formula (\ref{eqn:variationalsuffstat}), the marginal likelihood of the data $D$ satisfy the following lower bound:
$$
m(D) = \int_{\mathcal{R}^d} p(\bbeta) \ p(D \mid \bbeta)  \ d \beta  \geq  \int_{\mathcal{R}^d} p(\bbeta) \  \prod_{t=1}^N \exp\{f_t(\bbeta, \xi_t) \} \ d \bbeta \, .
$$
The second integral is analytically available, owing to the fact that the lower bound $f_t(\bbeta, \xi_t)$ from (\ref{eqn:variationalsuffstat}) is a quadratic form in $\bbeta$.  The variational-Bayes approach is to choose $\xi_t$ to make this lower bound on the marginal likelihood as tight as possible.  It leads to $\xi_t^2(\bbeta) =  x_t^T V_{\bbeta} x_t + (x_t m_{\bbeta})^2$, where $(m_\bbeta, V_\bbeta)$ are the mean and variance of the previous approximation, given in (\ref{eqn:completedataposterior}).  By comparison, the EM algorithm implicitly sets $\xi_t(\bbeta) = x_t^T \bbeta = x_t^T m_{\bbeta}$.  Notice that they differ by a factor that involves the Mahalonobis norm of $x_t$.

Thus we would summarize the key similarities and differences between the approaches as follows.  First, both algorithms yield updates for $\hat{\bbeta}(\lambda_1, \ldots, \lambda_N)$ that are identical in their functional dependence upon ``local'' parameters $\lambda_t$, even though these local parameters have different interpretations.  Moreover, both algorithms yield updates $\hat{\lambda}_t (\cdot)$ that are identical in functional form.  But the EM algorithm treats $\lambda_t$ as missing data having a known conditional expected value, and therefore evaluates $\hat{\lambda}_t (\cdot)$ at the current value of the linear predictor.  In contrast, the variational-Bayes approach, $\hat{\lambda}_t (\cdot)$ is treated as a function of a further variational parameter $\xi_t$, with the $\xi_t$ chosen to maximize the lower bound on the marginal likelihood of the data.  Thus at each step the chosen $\lambda_t$'s are different because of the different arguments at which $\hat{\lambda}_t (\cdot)$ is evaluated.

Finally, both algorithms yield an approximate Gaussian posterior that can be interpreted as a complete-data posterior distribution in an EM, where the imputed data have a Polya-Gamma conditional distribution.  In the EM approach, this Gaussian approximation is centered at the posterior mode.  In the variational-Bayes approach, it is centered on some other point in the parameter space that, by definition, has a lower posterior density.  By a parallel line of reasoning, one also concludes that the stationary point of EM algorithm leads to a worse lower bound on the marginal likelihood, although it is not clear how the tightness of this lower bound translates into accuracy in approximating the posterior for $\bbeta$.  In our numerical experiment reported in the supplement (which used MCMC as a benchmark), the variational posterior was typically centered somewhere between the mean and mode of the true posterior distribution (which is often quite skewed, especially for larger coefficients).  The posterior variances of the two algorithms tend to be nearly identical, and much smaller than the true variances.  In fact, we can identify the degree of overconfidence of both methods using standard tools from EM theory, specifically the fraction of missing information introduced by the PG data augmentation scheme.  See \cite{dempster:laird:rubin:1977} and \cite{louis:1982}.

\section{Extensions}

\subsection{Sparsity}

We can also introduce a penalty function $P(\beta)$ to the expected log likelihood.  The obvious and popular choice is the $\ell^1$ penalty, leading to a new complete-data objective function:
$$
\tilde{Q}_{\lambda}(\bbeta) = -\frac{1}{2} \bbeta^T S \bbeta + \beta^T d - \lambda \sum_{j=1}^p |\beta_j | \, .
$$
This is a convex problem that can be solved efficiently using coordinate descent and an active-set strategy \cite{balakrishnan:madigan:2008}.  Most updates are of order $k^2$, where $k$ is the size of the active set; full $d^2$ steps need be taken only rarely.  The updates can be made still more efficient by exploiting any sparsity that may be present in the design matrix $X$.

There are many proposals for fitting sparse logistic regression models; see \cite{fried:hastie:tibs:2010} for a review.  Most of these methods share one thing in common: they treat the likelihood $L(\bbeta)$ using the same device employed in Fisher scoring, or iteratively re-weighted least squares (IRLS).   That is, given the current estimates of the parameters $\tilde{\bbeta}$, a quadratic approximation to the log-likelihood is formed:
\begin{equation}\label{quadratic approximation}
l_Q(\bbeta) = -\frac{1}{2}\sum_{t=1}^Nw_t(z_t-x_t^T\bbeta)^2+C(\tilde{\bbeta})^2 \, ,
\end{equation}
where $C$ is constant in the parameter, and where
\begin{eqnarray}\label{eqn:WLS weights}
z_t &=& x_t^T\bbeta + \frac{y_t - \tilde{p}(x_t)}{\tilde{p}(x_t)(1-\tilde{p}(x_t))} \quad \quad \mbox{(working responses)}\\
w_t &=& \tilde{p}(x_t)(1-\tilde{p}(x_t)) \quad \quad \mbox{(weights)} \, ,
\end{eqnarray}
with the estimated success probability $\tilde{p}(x_t)$ evaluated at the current parameters.  The various approaches differ in the manner by which one finds the solution to the penalized, iteratively re-weighted least squares objective function:
$$
\min_{\bbeta \in R^p} \{ -l_Q(\bbeta) + \lambda P(\bbeta)\} \, ,
$$
where $P(\bbeta)$ is the penalty function.  As a result, these methods potentially inherit the numerical instability of iteratively re-weighted least squares, especially if initialized poorly.  In contrast, our approach handles the likelihood term using a missing-data argument, leading to a different quadratic form at each iteration.  Numerical evidence presented in the supplement suggests that the data-augmentation method is more robust, and leads to solution paths in $\lambda$ that achieve lower classification error than those based on penalized iteratively re-weighted least squares.

\subsection{Online EM}

\begin{algorithm}[t]
\begin{algorithmic}
\State Learning rate $c \in (0.5, 1)$; starting values $\beta^{(0)}$, $S_0$, $d_0$
\For{$t = 1, 2 \ldots$}
\State Read in batch of data $(m_{t}, y_t, X_t)$  of size $N$.
\State $\kappa_t \leftarrow (y_{t1} - m_{t1}/2, \ldots, y_{tN} - m_{tN}/2)^T$; $\psi_t \leftarrow X_t \beta^{(t-1)}$; $\gamma_t \leftarrow (t+1)^{-c}$
\For{$i = 1, \ldots, N$} $\omega_{ti} \leftarrow \frac{m_{ti}}{2 \psi_{ti}} \tanh(\psi_{ti}/2)$
\EndFor
\State $\Omega_t \leftarrow \mbox{diag}(\omega_{t1}, \ldots, \omega_{tM})$;  $S_t \leftarrow (1-\gamma_t) S_{t-1} +  \gamma_t X_t^T \Omega X_t$;  $d_t \leftarrow (1-\gamma_t) d_{t-1} + \gamma_t X_t \kappa_t$
\State $\beta \leftarrow \texttt{solve}(S_t, d_t)$
\EndFor
\end{algorithmic}
\caption{Online EM for logistic regression.}
\label{alg:ig2}
\end{algorithm}

We now consider an online version of the EM algorithm, which is much more scalable, has essentially the same convergence guarantees as batch EM, and operates without ever loading the whole data set into memory.  A good reference for the general theory of online EM algorithms is \cite{cappe:moulines:2009}.

Suppose we have observed up through data point $t$, and that we have a current estimate of the complete-data sufficient statistics; call these $d_t$ and $S_t$, in which case $\bbeta_t$ is the maximizer of the corresponding complete-data objective function (\ref{eqn:completedataposterior}).  Now we see a new triplet $(m_{t+1}, y_{t+1}, x_{t+1})$.  In the online E-step, we update the sufficient statistics as a convex combination of the old statistics and a contribution from the new data:
\begin{eqnarray*}
S_{t+1} &=& (1-\gamma_{t+1}) S_t + \gamma_{t+1} \hat{\omega}_t x_{t+1} x_{t+1}^T \\
d_{t+1} &=& (1-\gamma_{t+1}) d_t + \gamma_{t+1} x_{t+1} \kappa_{t+1} \, .
\end{eqnarray*}
The simplicity of this step owes to the linearity of the complete-data likelihood in $\omega_t$.  Here $\kappa_{t+1} = y_{t+1} - m_{t+1}/2$, and $\hat{\omega}_{t+1}$ is given by Formula (\ref{eqn:omegahat2}), evaluated at the current estimate $\bbeta_t$.  Then in the $M$-step, we solve (or take a step towards the solution of) the log-Gaussian posterior density with sufficient statistics $S_{t+1}$ and $d_{t+1}$.

This can also be done in mini-batches of size $N$, with the obvious modification to the the sufficient-stat updates from the single-data-point case.  Let $m_t$ and $y_t$ be $N$-vectors of trials and successes for the observations in batch $t$, and $X_t$ as an $N \times p$ matrix of regressors for these $N$ observations.  The updates are then
\begin{eqnarray*}
S_{t+1} &=& (1-\gamma_{t+1}) S_t + \gamma_{t+1}  X_{t+1}^T \Omega_{t+1} X_{t+1} \\
d_{t+1} &=& (1-\gamma_{t+1}) d_t + \gamma_{t+1} X_{t+1} \kappa_{t+1} \, ,
\end{eqnarray*}
where $\Omega_{t+1}$ is the diagonal matrix of $\omega$ terms arising from the current mini-batch, each computed using Equation  \ref{eqn:omegahat2} evaluated at the current estimate for $\bbeta$.  Empirically, we have found that mini-batch sizes within an order of magnitude of $d$ (the dimension of $\bbeta$) have given robust performance, and that the method is unstable when processing only a data point at a time.  

  In our experience, the online version of the algorithm is usually much faster than the batch algorithm, even in batch settings.  It is also just as accurate. The only tuning parameter is the learning rate $\gamma_t$, which arises in all online-learning algorithms.  The formal convergence requirement from \cite{cappe:moulines:2009} is that the $\gamma_t$ follow a decay schedule where
$$
\sum_{t=1}^{\infty} \gamma_t = \infty \quad \mbox{and} \quad \sum_{t=1}^{\infty} \gamma_t^2 < \infty \, .
$$
Thus we adapt to new data fast enough (the non-summability condition), but not so fast that the estimate bounces around the truth forever without ever converging (the square-summability condition).  This is the usual requirement of the updating rule in any stochastic-approximation algorithm.  A simple way of ensuring this is to choose $\gamma_t \propto (t + t_0)^{-c}$ for some $c \in (0.5,1)$.

We follow the advice of  \cite{cappe:moulines:2009}, which is to use $c$ very close to $0.5$, and then to smooth the estimate of $\bbeta$ using Polyak-Ruppert averaging: taking the final estimate of $\bbeta$ to be the average solution for $\bbeta$ over the final $T-K$ iterations, where $K$ is large enough so that the algorithm has settled down.

\begin{figure}[t]
\begin{center}
\includegraphics[width=0.9\textwidth]{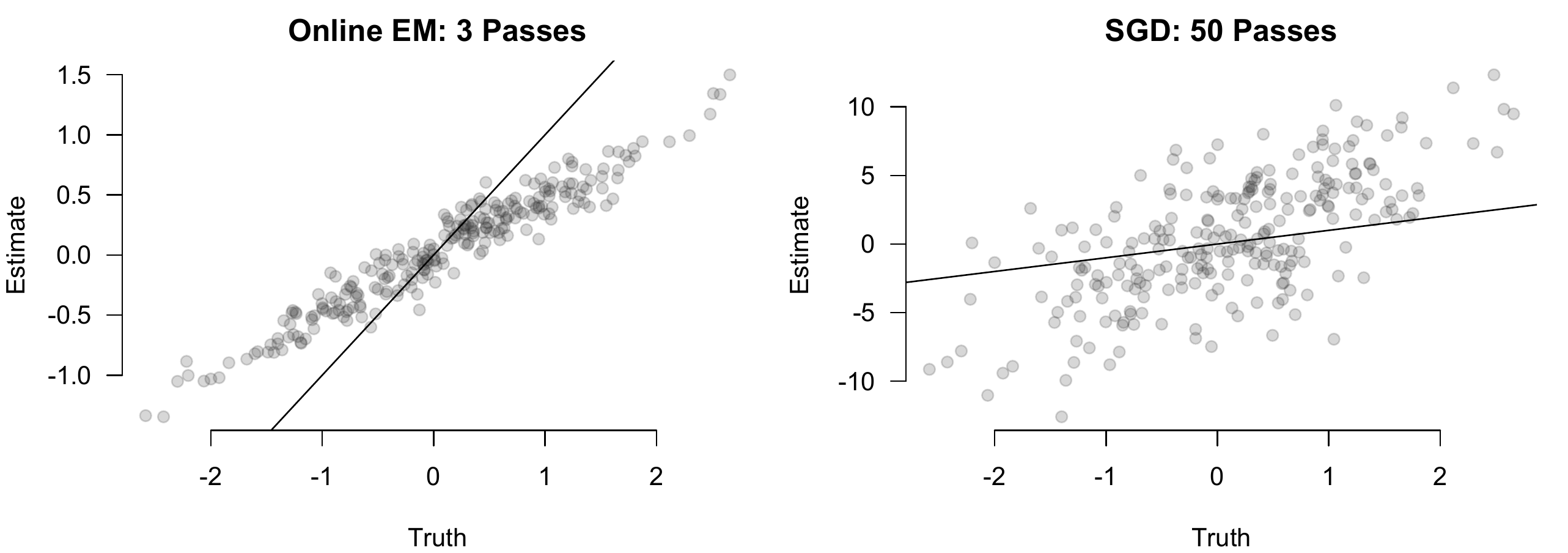}
\caption{\label{fig:onlineEM} Performance of the online EM algorithm (3 passes, no Polyak-Ruppert averaging) versus stochastic gradient descent (50 passes) for a logistic regression problem with correlated predictors. }
\end{center}
\end{figure}

The online EM is a second-order method that requires working with a $d \times d$ matrix.  We therefore do not anticipate that it will scale to the very largest of problems, at least with present computing technology.  Indeed, for ultra-large-scale problems such as those considered by \cite{langford:li:zhang:2009}, of order $d \approx 10^9$, it is infeasible to even form the sufficient statistics $S$ and $d$, much less to solve the linear system in the final line of Algorithm \ref{alg:ig2}.  Here methods based on stochastic gradient descent  (e.g.~\cite{littlestone:etal:1995}) are probably the better choice, if only because nothing else beyond a first-order method will run.

Nonetheless, there is a vast middle ground of potential problems where online EM may offer notable advantages compared to stochastic gradient-descent: problems that are too large to be solved easily by batch methods, but that are ``small enough'' (which may still mean thousands of parameters) to make forming the sufficient statistics $S$ and $d$ feasible.  The fundamental tradeoff is one of per-iteration cost (where a first-order method clearly wins) versus distance traversed per iteration (where a second-order method clearly wins).  When the design points are highly collinear, SGD can be extremely slow to converge, and the per-iteration cost of online EM may be worth it.   As argued by \cite{bottou:bousquet:2008}, constants matter when comparing the actual efficiency of first-order versus second-order methods, and these constants involve features of the the $x_t$'s that may not be known beforehand.

As an example, Figure \ref{fig:onlineEM} shows the performance of the online EM on a simulated data set of 250 predictors and 100,000 observations with collinear predictors.  We benchmarked against stochastic gradient descent.  The design points $x_t$ were correlated multivariate normal draws, with covariance matrix $\Sigma = B B^T + 0.1I$, $B$ being a $250 \times 50$ factor loadings matrix with standard normal entries.  The columns of $X$ were subsequently rescaled to have marginal variance $1/p$, ensuring a standard normal distribution for the linear predictors $\psi_t$.  We used multiple passes for both algorithms (50 passes for SGD, 3 passes for online EM), with each pass scanning the data points in a random order.  The algorithms used a similar decay schedule, with the online EM processing data in batches of size 500.  We chose 3 and 50 to yield similar computing times for the two methods.  Although this ratio will clearly depend upon the size of the problem, at least for this (nontrivially large) problem, online EM is clearly the more efficient choice.

There are three other advantages of online EM, at least for problems that live in this middle ground.  First and most obviously, online EM gives an approximation to the entire posterior distribution, rather than just a point estimate (assuming that the conditional sufficient statistics are re-scaled by the total number of data points processed).  Even if the error bars that arise from the complete-data posterior are too optimistic (ala variational Bayes), they are better than nothing, and still allow meaningful relative comparisons of uncertainty.  Second, stochastic gradient descent is notoriously brittle with respect to the decay schedule of the learning rate, with poor choices leading to outrageously bad performance.  Online EM, especially when coupled with Polyak-Ruppert averaging, tends to be much more robust.

Finally, merging SGD with sparsity-promoting penalties is known to be challenging, and is typically restricted to an $\ell^1$ penalty \cite{langford:li:zhang:2009}.  In contrast, nonconvex (heavy-tailed) penalties are easily incorporated into both the batch and online versions of our approach.  For example, the authors of \cite{balakrishnan:madigan:2008} report considerable success with a sparse second-order method even for very large problems.  The disadvantage of their method is that convergence is not formally guaranteed even without sparsity; by contrast, the convergence of our approach follows straightforwardly from standard results about online EM.

\section{Remarks}

We have introduced a family of expectation-maximization algorithms for binomial and negative-binomial regression.  The existence of such an algorithm, and its intimate connection with variational Bayes, have not previously been appreciated in the machine-learning community.  Indeed, the fact that the same local parameters $\lambda_t$ arise in both algorithms, despite having very different interpretations and constructions, is noteworthy and a bit puzzling.  It suggests interesting connections between two fundamental operations in statistics---namely profiling and marginalization---that are not usually thought of as being part of the same constellation of ideas.  The strengths of our method are: (1) that it is very robust thanks to the ascent property of EMs; (2) that it leads to error bars very similar to those that arise from variational Bayes, but with a guarantee of consistency; (3) that is can easily be extended to incorporate sparsity-inducing priors; and (4) that it leads straightforwardly to a robust second-order online algorithm whose convergence properties are easily established.  Further research is clearly needed on the performance of the online EM, in order to establish the circumstances (as suggested by Figure \ref{fig:onlineEM}) in which it will outperform stochastic gradient descent for a fixed computational budget.

In conclusion, we refer interested readers to the appendix, which contains several details not described here:
\begin{itemize}
\item A comparison of the quality of posterior approximations arrived at by VB, EM, and QNEM on a simple example.
\item A study of the data-augmentation approach to handling logit likelihoods, versus penalized iteratively re-weighted least squares, in the context of sparsity-inducing priors.
\item An extension to the multinomial case.
\end{itemize}

\singlespace
\bibliographystyle{unsrt}
\bibliography{masterbib}

\appendix

\section{Variational Bayes, EM, and QNEM}

This section describes a simple simulate example whose purpose is to understand two questions. (1) How similar are the variational-Bayes approximation and the complete-data log posterior arising from the EM algorithm?  (2) Does using the estimated ``remainder'' Hessian matrix from QNEM lead to more sensible error bars, when judged against the posterior distribution calculated by MCMC?

We simulated a small data set with $d=10$ coefficients $\beta = (-3, -2.33, -1.67, \ldots, 2.33, 3)$, independent, standard normal predictors $x_{tj}$, and $n=250$ observations from a logistic-regression model.  We then used four algorithms to estimate $\bbeta$:
\begin{itemize}
\item Markov Chain Monte Carlo, using the method from \cite{polson:scott:windle:2012a} and the R package \verb|BayesLogit| \cite{bayeslogit:2013}.  This algorithm results in a uniformly ergodic Markov chain with well understood convergence properties, and so is a reasonable gold standard.
\item Expectation-maximization, as described in the main manuscript.
\item Expectation-maximization with quasi-Newton acceleration.
\item Variational Bayes, as described in the main manuscript and in \cite{jaakkola:jordan:2000}.
\end{itemize}
In all cases a vague mean-zero normal prior with precision $10^{-5}$ on each coefficient was used.

\begin{figure}
\begin{center}
\includegraphics[width=\textwidth]{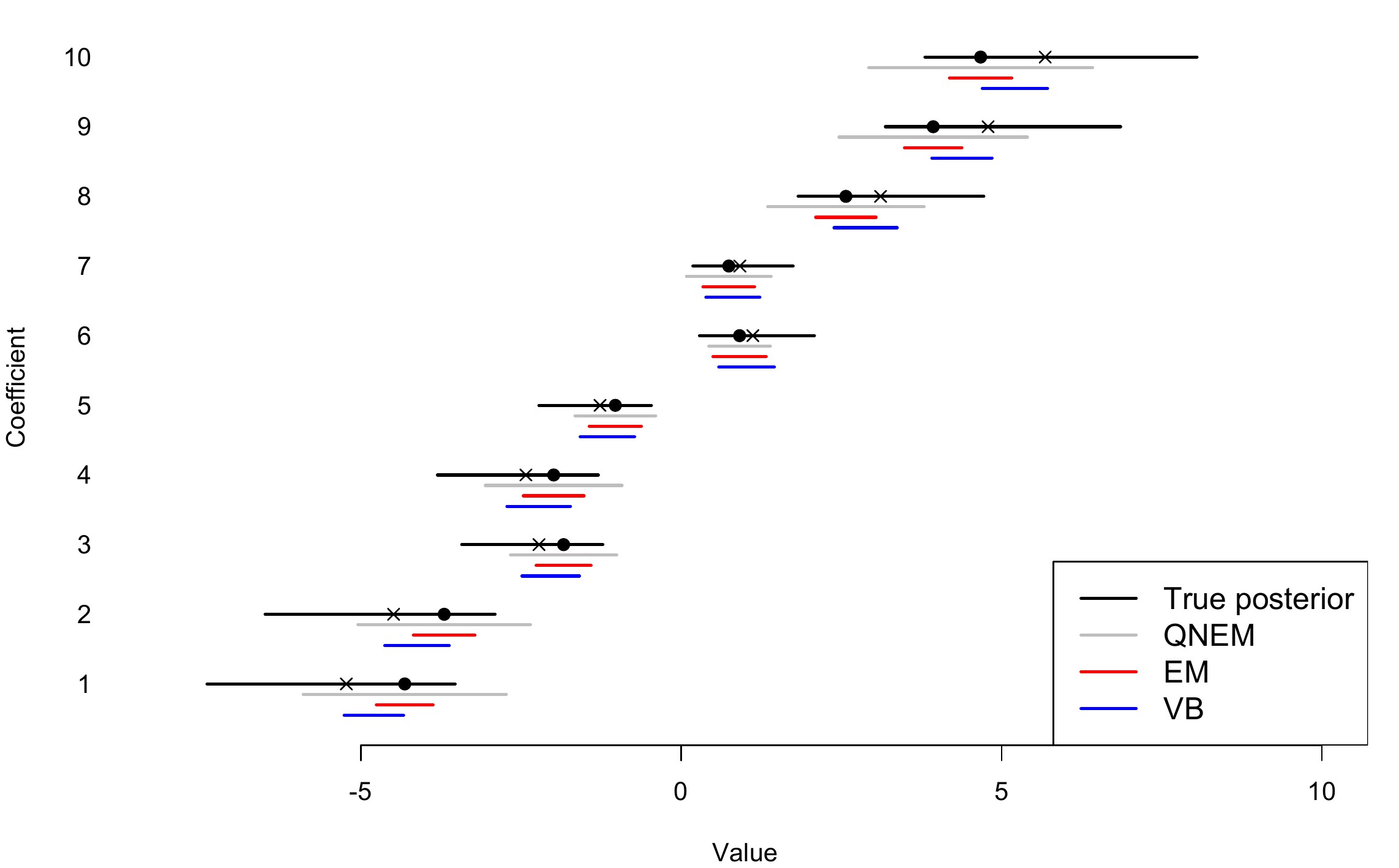}
\caption{\label{fig:vbqnem} Results on the simulated data set.  The black lines are the 95\% central credible intervals from the MCMC sample.  Black dots: posterior modes.  Black crosses: posterior means.  The other lines are approximate $95\%$ credible intervals for the other three methods.}
\end{center}
\end{figure}

Using $10^6$ MCMC samples (after an initial burnin of $10^5$ samples), we computed central $95\%$ credible intervals for each coefficient.  These are show as black lines in Figure \ref{fig:vbqnem}.  The posterior means are shown as black crosses, and the posterior modes as black dots (the marginal posteriors for this problem are notably skewed).  Using the other three methods, we constructed symmetric $95\%$ approximate credible intervals using the estimate and the approximate posterior standard deviations.  As Figure 1 shows, the VB and EM estimates have essentially identical approximate posterior standard deviations.  The only visible difference is that the EM estimates are centered about the posterior mode, while the VB estimates are centered somewhere between the mean and the mode.   The QNEM estimates have significantly higher spread than either the VB or EM intervals, and in this respect are much more in line with the true standard deviations.

\section{Sparsity in logistic regression}

\subsection{Penalized IRLS}

The main manuscript makes the following claim:
\begin{quotation}
[Existing methods for sparse logistic regression] potentially inherit the numerical instability of iteratively re-weighted least squares, especially if initialized poorly.  In contrast, our approach handles the likelihood term using a missing-data argument, leading to a different quadratic form at each iteration.  Numerical evidence presented in the supplement suggests that our method is more robust, and leads to solution paths in $\lambda$ that achieve lower classification error than those based on penalized iteratively re-weighted least squares.
\end{quotation}

The goal of this section is to provide the evidence in support of this claim.  Consider a quadratic approximation to the log-likelihood:
\begin{equation}\label{quadratic approximation}
l_Q(\bbeta) = -\frac{1}{2}\sum_{t=1}^Nw_t(z_t-x_t^T\bbeta)^2+C(\tilde{\bbeta})^2 \, ,
\end{equation}
where $C$ is constant in the parameter, and where
\begin{eqnarray}\label{eqn:WLS weights}
z_t &=& x_t^T\bbeta + \frac{y_t - \tilde{p}(x_t)}{\tilde{p}(x_t)(1-\tilde{p}(x_t))} \quad \quad \mbox{(working responses)}\\
w_t &=& \tilde{p}(x_t)(1-\tilde{p}(x_t)) \quad \quad \mbox{(weights)} \, ,
\end{eqnarray}
with the estimated success probability $\tilde{p}(x_t)$ evaluated at the current parameters.  Most existing approaches use this approximation to the likelihood, and differ in the manner by which one finds the solution to the penalized, iteratively re-weighted least squares objective function:
$$
\min_{\bbeta \in R^p} \{ -l_Q(\bbeta) + \lambda P(\bbeta)\} \, ,
$$
where $P(\bbeta)$ is the penalty function.  In this section, we simply consider the case $n_t = 1$ for all $t$, and the $\ell^1$ penalty: $P(\bbeta) = \sum_{j=1}^P |\beta_j|$

For example, in \cite{fried:hastie:tibs:2010}, a coordinate descent algorithm is used to solve the penalized weighted least-squares problem.  The coordinate-wise update has the form
\begin{equation}\label{update betaj}
\tilde{\beta_j} \leftarrow \frac{S(\sum_{i=1}^Nw_ix_{ij}(y_i-\tilde{y_i^{(j)}}),\lambda)}{\sum_{i=1}^Nw_ix_{ij}^2}
\end{equation}
where $\tilde{y_i^{(j)}} = \bm{x_i}^t\tilde{\bbeta}-x_{ij}\beta_j$ is the fitted value excluding the contribution from $x_{ij}$, and $S(z,\gamma)$ is the soft-thresholding operator with value
\[
 sign(z)(|z|-\gamma)_+ =
  \begin{cases}
   z-\gamma & \text{if } z>0 \quad \text{and} \quad \gamma < |z| \\
   z+\gamma & \text{if } z<0 \quad \text{and} \quad \gamma <|z|\\
   0       & \text{if } \gamma \geq |z|
  \end{cases}
\]

Algorithm \ref{alg:WLS_CD} summarizes the approach for fixed $\lambda$.
\begin{algorithm}[t]
\begin{algorithmic}
\State Data: $(y_t)$ for $t = 1, \ldots, N$; design matrix $X$ having $x_t^T$ as row $t$.
\State Starting value: $\beta$
\Repeat 
\State Update the quadratic approximation $l_Q$ using current $\bbeta$ as in \ref{quadratic approximation}
\For{$j = 1, \ldots, P$}
\State Update $\beta_j$ using coordinate descent as in \ref{update betaj} 
\EndFor
\Until $\beta$ converges.
\end{algorithmic}
\caption{Weighted Least Square updates with coordinate descent.}
\label{alg:WLS_CD}
\end{algorithm}

Our approach, on the other hand, is to employ the Polya-Gamma data augmentation trick for handling the logit likelihood and to represent the prior distribution as a mixture of normals:
$$
p(\beta_j|\lambda) = \int_0^{\infty} \phi(\beta_j|0,\gamma_j/\lambda^2)dP(\gamma_j) \, .
$$
We then use the linear conjugate gradient algorithm to optimize the resulting log posterior distribution.  The complete-data log posterior, conditioning on the ''missing data'' $\omega_t, \gamma_t$, can be computed as
$$
\tilde{Q}_{\lambda}(\bbeta) = -\frac{1}{2} \bbeta^T S \bbeta + \beta^T d - -\frac{1}{2}\lambda^2\sum_{j=1}^P\gamma_j^{-1}\beta_j^2 \, ,
$$
In M step, we employ the same step as in Batch EM without sparsity. The complete-data sufficient statistics are
\begin{eqnarray*}
S &=& X^T\Omega X+\lambda^2\Gamma^{-1}\\
d &=& X^T\kappa
\end{eqnarray*}
Thus $\hat{\bbeta}$ solves the linear system $S\hat{\bbeta} = d$.

To perform the E-step, we can simply replace $\gamma_j$ and $\omega_t$ with their conditional expectations $\gamma_j^{(g)}$ and $\omega_t^{(g)}$, given the observed data and the current $\bbeta^{(g)}$.  The conditional moments $\hat{\gamma_j^{(g)}} = E(\gamma_j|\bbeta^{(g)},y)$ and $\hat{\omega_j^{(g)}} = E(\omega_j|\bbeta^{(g)},y)$ are given by the following expressions:
\begin{eqnarray}
\hat{\gamma_j} &=& \lambda |\beta_j| \\
\hat{\omega_t} &=& \left( \frac{1}{2x_t^T\bbeta^{(g)}} \right) \tanh(x_t^T\bbeta^{(g)}/2)
\end{eqnarray}

We can solve the linear system $S\hat{\bbeta}=d$ directly at each step, or we can instead use the linear conjugate gradient algorithm to solve the system.  Here we have different variations for conjugate gradient algorithm. We can try n-step ($n \leq p$)Conjugate Gradient (CG) Method, or $\epsilon$-tolerance CG Method, or just one-step CG method. In p-step CG method, we force the algorithm to run until it reached the exact solution.  In $n, (n < p)$-step CG method, we stop after $n$ iterations, while in $\epsilon$-tolerance CG Method we stop the algorithm when $x$ converges in $\epsilon$-tolerance.

\begin{algorithm}[t]
\begin{algorithmic}
\State Data: $(y_t)$ for $t = 1, \ldots, N$; design matrix $X$ having $x_t^T$ as row $t$.
\State $\kappa \leftarrow (y_1 - 1/2, \ldots, y_N - 1/2)^T$
\State $d \leftarrow X^T \kappa$
\State Starting value: $\beta$
\Repeat 
\For{$t = 1, \ldots, N$}
\State $\psi_t \leftarrow x_t^T \beta$
\State $\omega_t \leftarrow \frac{n_t}{2 \psi_t} \tanh(\psi_t/2)$
\EndFor
\For{$j = 1, \ldots, P$}
\State $\gamma_t \leftarrow \lambda |\beta_j|$
\EndFor
\State $\Omega \leftarrow \mbox{diag}(\omega_1, \ldots, \omega_N)$
\State $\Gamma^{-1} \leftarrow \mbox{diag}(\gamma_1^{-1},\ldots, \gamma_P^{-1})$
\State $S \leftarrow X^T \Omega X + \lambda^2\Gamma^{-1}$
\State $\beta \leftarrow S^{-1} d$ Solve the system by conjugate gradient
\Until $\beta$ converges.
\end{algorithmic}
\caption{Batch EM with lasso prior for logistic regression.}
\label{alg:ig3}
\end{algorithm}

\begin{algorithm}[t]
\caption{$\epsilon$-tolerance CG Method}
\begin{algorithmic}
\State $i\Leftarrow 0$
\State $r\Leftarrow d-Sx$
\State $b\Leftarrow r$
\State $\delta_{new} \Leftarrow r^T r$
\State $\delta_0 \Leftarrow \delta_{new}$
\Repeat 
\State $q \Leftarrow Sb$
\State $\alpha \Leftarrow \frac{\delta_{new}}{b^T q} $
\State $x \Leftarrow x + \alpha b$
\State $r \Leftarrow r - \alpha q$
\State $\delta_{old} \Leftarrow \delta_{new}$
\State $\delta_{new} \Leftarrow r^T r$
\State $\beta \Leftarrow \frac{\delta_{new}}{\delta_{old}}$
\State $b \Leftarrow r + \beta b$
\State $i \Leftarrow i+1$
\Until $i<i_{max}$ and $\delta_{new} > \epsilon^2 \delta_0$ 
\end{algorithmic}
\label{alg:CG}
\end{algorithm}

\subsection{Comparison}
Notice that our data-augmentation trick can also be considered as a re-weighted least squares algorithm, but with different weights. The corresponding weights and working responses are
\begin{eqnarray}\label{eqn:aug weights}
\omega_t = \left( \frac{1}{2x_t^T\bbeta^{(g)}} \right) \tanh(x_t^T\bbeta^{(g)}/2) \quad \mbox{(weights) }\\
z_t = \frac{2y_t - 1}{\omega_t} \quad \mbox{(working responses)}
\end{eqnarray}

Here we will test different combinations of weights derived from IRLS (1) versus our data-augmentation scheme (2); and coordinate descent method (3) versus conjugate gradient method (4) for solving each iteration's subproblem.  There are four combinations, and our goal is to see which combination performs best.  The numerical results are presented in the next sub-section.

%
%

\subsection{Batch EM with Bridge Penalty}
Data augmentation also allows us to deal with bridge penalty/prior as well. Here we want to minimize
\begin{equation}
l(\bbeta) =  \sum_{t=1}^N (y_t \log p(x_t)- (n_t-y_t)\log(1-p(x_t)))- \lambda \sum_{j=1}^P|\beta_j|^{\alpha}
\end{equation}
with $0<\alpha<1$.  The bridge penalty can be represented as a mixture of normals: 
$$
p(\beta_j) = \int \phi(\beta_j;0,\gamma_j/\lambda^2)d\gamma_j
$$
where $p(\gamma_j)$ follows a Stable distribution.  The conditional moments $\hat{\gamma_j^{-1(g)}} = E(\gamma_j^{-1}|,\bbeta^{(g)},y)$ are given by
\begin{equation}
\hat{\gamma_j^{-1(g)}} = \alpha \beta_j^{\alpha-2} \mbox{sign}(\beta_j)/\lambda^2
\end{equation}

The corresponding algorithm for our batch EM is show above (Algorithm \ref{alg:WLS_CG_bridge}).

\begin{algorithm}[t]
\begin{algorithmic}
\State Data: $(y_t)$ for $t = 1, \ldots, N$; design matrix $X$ having $x_t^T$ as row $t$.
\State $\kappa \leftarrow (y_1 - 1/2, \ldots, y_N - 1/2)^T$
\State $d \leftarrow X^T \kappa$
\State Starting value: $\beta$
\Repeat 
\For{$t = 1, \ldots, N$}
\State $\psi_t \leftarrow x_t^T \beta$
\State $\omega_t \leftarrow \frac{n_t}{2 \psi_t} \tanh(\psi_t/2)$
\EndFor
\For{$j = 1, \ldots, P$}
\State $\gamma_j \leftarrow \alpha \beta_j^{\alpha-2} \mbox{sign}(\beta_j)/\lambda^2$
\EndFor
\State $\Omega \leftarrow \mbox{diag}(\omega_1, \ldots, \omega_N)$
\State $\Gamma^{-1} \leftarrow \mbox{diag}(\gamma_1^{-1},\ldots, \gamma_P^{-1})$
\State $S \leftarrow X^T \Omega X + \lambda^2\Gamma^{-1}$
\State $\beta \leftarrow \mbox{Solve}(S, d)$ by conjugate gradient
\Until $\beta$ converges.
\end{algorithmic}
\caption{Batch EM with bridge prior for logistic regression.}
\label{alg:WLS_CG_bridge}
\end{algorithm}

\subsection{Numerical Results}

\begin{figure}
\begin{center}
\includegraphics[width=0.8\textwidth]{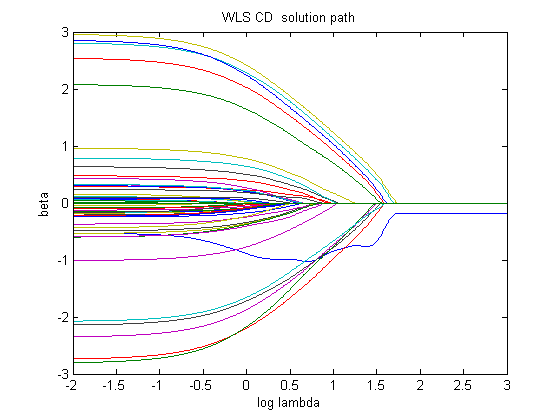}
\caption{\label{fig:beta1-sm-sl} Solution path for algorithm IRLS+CD. The x axis is $log_{10} \lambda$.}
\end{center}
\end{figure}

\begin{figure}
\begin{center}
\includegraphics[width=0.8\textwidth]{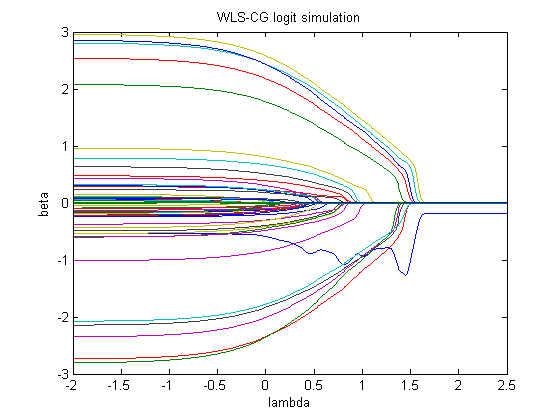}
\caption{\label{fig:beta2-sm-sl} Solution path for algorithm IRLS+CG. The x axis is $log_{10} \lambda$.}
\end{center}
\end{figure}

\begin{figure}
\begin{center}
\includegraphics[width=0.8\textwidth]{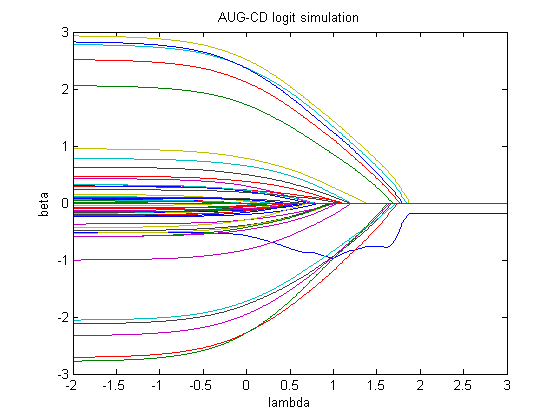}
\caption{\label{fig:beta3-sm-sl} Solution path for algorithm Data Augmentation+CD. The x axis is $log_{10} \lambda$.}
\end{center}
\end{figure}

\begin{figure}
\begin{center}
\includegraphics[width=0.8\textwidth]{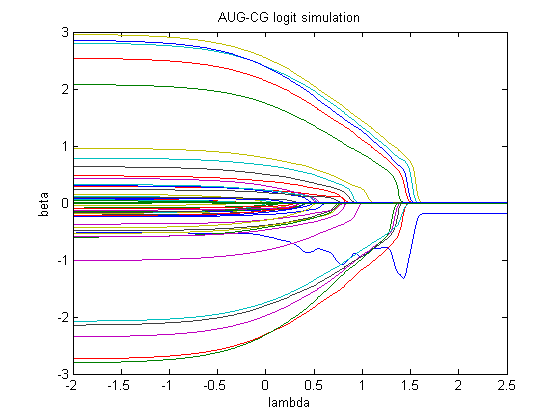}
\caption{\label{fig:beta4-sm-sl} Solution path for algorithm Data Augmentation+CG. The x axis is $log_{10} \lambda$.}
\end{center}
\end{figure}

We now describe our numerical experiments on simulated data sets.  For each dataset, we model them as logistic regression problems. We tested the four combinations of WLS/DA (for the likelihood) and CG/CD (for the prior/penalty) stated above for lasso penalty.  We also tested our data augmentation approach combining (using the conjugate gradient algorithm)assuming a bridge penalty with $\alpha$ equal both to 0.5 and 0.75. 

We simulated $X$ as a $500$ by $50$ matrix with random $0$ and $1$ elements, and set the true $\beta$ equal to $\sqrt{5}$ (with alternating signs) for the first 10 coefficients, with the rest being zero.  The four graphs of the solution paths  (as a function of $\lambda$) for each algorithm are show in Figures \ref{fig:beta1-sm-sl} through \ref{fig:beta4-sm-sl}. The grid size is 0.01.  There are clear differences among the solutions paths for each algorithm, with data augmentation algorithms leading to systematically lower values of the penalized likelihood.

\begin{figure}
\begin{center}
\includegraphics[width=0.8\textwidth]{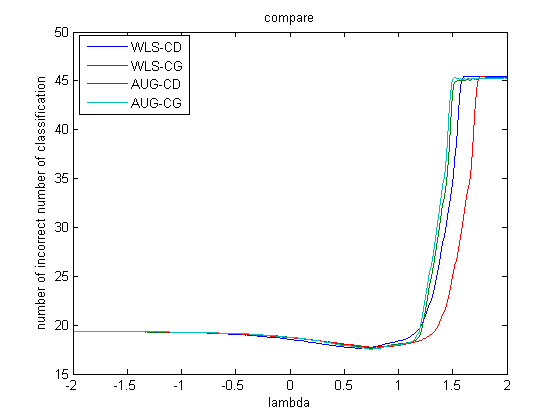}
\caption{\label{fig:nic-sm-4} The figure compares the number of average incorrect number of classification for all four algorithms along $log_{10}\lambda$.}
\end{center}
\end{figure}

To get a further idea of which algorithm is providing the best solution, we also looked at out of sample performance.  Specifically, we used 80\% of the observations to estimate $\beta$ for each value of $\lambda$ using each algorithm, and then used the estimated $\beta$ to predict the remaining 20\% observations. We calculated the mean of the incorrect classifications $y_t = \{0,1\} $ across over 1000 random train/test splits.  The results as a function of $\lambda$ are shown in Figure \ref{fig:nic-sm-4}.  Overall the DA + coordinate-descent combination is a bit better than the others.

\section{Batch EM for multinomial regression}

\subsection{Penalized Partial IRLS}

When the response variable $y$ has more than $2$ levels, the linear logistic regression model can be generalized to a multinomial logistic model. Suppose we observe data $(y_1,\bm{x_1},\ldots,(y_N,\bm{x_N}))$, where $y_t \in \{1,\ldots K\}$ is an integer outcome denoting membership in one of $K$ classes, and $\bm{x_t}$ is a $p$-vector of predictors. Under the multinomial logit model, the probability of observation $y_t$ falling in class $k$ is assumed to be
\begin{equation}\label{multinomial prob}
\theta_{tk} = P(y_t = k) = \frac{\exp(\bm{x_i^T}\bbeta_k)}{\sum_{l=1}^K \exp (\bm{x_i^T}\bbeta_l)}
\end{equation}
where $\bbeta_k$ is a $p$-vector of regression coefficients for class $k$.

Let $\bm{Y}$ be the $N \times K$ indicator response matrix, with elements $Y_{tl} = I(y_t = l)$. Then we want to maximize the penalized log-likelihood:
\begin{equation}\label{mul penalized log likeli}
l(\bbeta) = \sum_{t=1}^N[\sum_{k=1}^K Y_{tk}log\theta_{tk}+(1-Y_{tk})log(1-\theta_{tk})]-\sum_{k=1}^K\lambda P(\bbeta_k)
\end{equation}
For lasso penalty, the penalty function is $P(\bbeta_k) = \sum_{j=1}^P|\beta_{jk}|$.

Allowing only $\bbeta_k$ to vary for a single class a time, a partial quadratic approximation given current estimate $\hat{\bbeta}$ to the log-likelihood part of (\ref{mul penalized log likeli}) is
\begin{equation}\label{m:quadratic app}
l_{Qk}(\bbeta_k) = -\sum_{t=1}^N w_{tk}(z_{tk}-\bm{x_t}^T\bbeta_k)^2+C(\{\hat{\bbeta}\})
\end{equation}
where
\begin{eqnarray}\label{eqn:m weight updates}
z_{tk} = \bm{x_t}^T\bbeta_k + \frac{y_{tk}-\hat{p}_k(x_t)}{\hat{p}_k(x_t)(1-\hat{p}_k(x_t))}\\
w_{tk} = \hat{p}_k(x_t)(1-\hat{p}_k(x_t)) \, .
\end{eqnarray}

The approach is similar to logistic regression, except now for each $\lambda$ the classes are cycled over in the outer loop, and for each class $k$ there is a partial quadratic approximation $l_Q(\bbeta_k)$ about the current parameters $\hat{\bbeta}$. Then coordinate descent is used to solve the penalized weighted least-squares problem.  Notice that $\{\bbeta_k\}_1^K$ and $\{ \bbeta - \bm{c}\}$ give the same log-likelihood, yet a different penalty. Therefore, the estimate $\{ \bbeta_k\}_1^K$ can be improved by solving
$$
\min_{c \in R^P} \sum_{k=1}^K P(\bbeta_k-c).
$$
This can be done separately for each coordinate, and leads to $\bbeta$ being recentralized by choosing $c_j$ to be the median of the $\beta_{jk}, 1\leq k \leq K$. 

For each $\lambda$, the algorithm is:

\begin{algorithm}
\begin{algorithmic}
\State Data: $(y_{tk})$ for $t = 1, \ldots, N$ and $k = 1, \ldots, K$; design matrix $X$ having $x_t^T$ as row $t$.
\State Starting value: $\beta$
\Repeat  
\State Cycle over $k \in \{ 1, 2, \ldots, K, 1, 2, \ldots\}$ 
\State Update the quadratic approximation $l_{Qk}$ using current $\bbeta$ as in \ref{m:quadratic app}
\For{$j = 1, \ldots, P$}
\State Update $\beta_j$ using coordinate descent as in \ref{eqn:m weight updates} 
\EndFor
\State After each cycle
\For{$j = 1, \ldots, P$}
\State $\bbeta_j = \bbeta_j - median(\bbeta_j)$ where $\bbeta_j = (\beta_{j1},\ldots, \beta_{jK})$
\EndFor
\Until $\beta$ converges.
\end{algorithmic}
\caption{multinomial regression with coordinate descent.}
\label{alg:m_WLS_CD}
\end{algorithm}

\subsection{Data augmentation and expectation/conditional maximization}

Our data augmentation approach can be used in a parallel fashion to the binomial logit case, leading to an ECM (expectation, conditional maximization) algorithm.  The difference is that we fix $\bbeta_1$ corresponding to class $1$ to be $(0,\ldots, 0)$ to ensure identifiability, and thus interpret the other coefficients in terms of changes in log-odds relative to the first category.  We phrase the problem as one of maximizing the posterior density
\begin{equation}
p(B|y)\propto \{\prod_{t=1}^N \prod_{k=1}^K \theta_{tk}^{y_{tk}}(1-\theta_{tk})^(1-y_{tk}) \}\cdot \exp\{-\sum_{k=2}^K\sum_{j=1}^P \lambda |\beta_j|\}
\end{equation}

Let $\eta_{tk} = \exp(\bm{x_t^T}\bbeta_k-c_{tk})/\{1+\exp(\bm{x_t^T}\bbeta_k-c_{tk})\}$, where $c_{tk}(\bbeta_{(-k)}) = \log \sum_{l \neq k}\exp(\bm{x_t^T}\bbeta_l)$. The conditional likelihood in $\bbeta_k$, given all other terms $\bbeta_{-k}$, can be written as
\begin{equation}
L(\bbeta_k|\bbeta_{-k},y) \propto \prod_{t=1}^N \{ \frac{\exp(\zeta_{tk}(\bm{x_t^T}\bbeta_k-c_{tk}))}{1+\exp(\zeta(\bm{x_t^T}\bbeta_k-c_{tk}))} \}
\end{equation}
where $\zeta_{tk}$ is the binary indicator $y_{tk}$ re-coded as $\pm 1$. Thus the conditional likelihood in $\bbeta_k$ looks like a logistic regression for the binary outcome $\zeta_{tk}$. And we can find a solution $\hat{B}$ by cycling through each block $\bbeta_k = (\beta_{k1},\ldots,\beta_{kp})^T$ of item-specific regression coefficients in turn, iterating this cycle until convergence. At each sub-step of the cycle, we are facing a logistic regression problem, which we have discussed in previous sections.  See Algorithm \ref{alg:multinomial AUG CG}.

\begin{algorithm}
\begin{algorithmic}
\State Data: $(y_{tk})$ for $t = 1, \ldots, N$ and $k = 1,\ldots,K$; design matrix $X$ having $x_t^T$ as row $t$.
\For{$k = 2,\ldots,K$}
\State $\kappa_k \leftarrow (y_{1k} - 1/2, \ldots, y_{1K} - 1/2)^T$
\State $d_k \leftarrow X^T \kappa_k$
\EndFor
\State Starting value: $\bbeta$
\Repeat 
\For{$k = 2,\ldots, K$}
\For{$t = 1, \ldots, N$}
\State $\psi_{tk} \leftarrow x_t^T \bbeta_k-\log \sum_{l \neq k} \exp(x_t^T \bbeta_l)$
\State $\omega_{tk} \leftarrow \frac{1}{2 \psi_{tk}} \tanh(\psi_{tk}/2)$
\EndFor
\For{$j = 1, \ldots, P$}
\State $\gamma_{jk} \leftarrow \lambda |\beta_{jk}|$
\EndFor
\State $\Omega_k \leftarrow \mbox{diag}(\omega_{1k}, \ldots, \omega_{Nk})$
\State $\Gamma_k^{-1} \leftarrow \mbox{diag}(\gamma_{1k}^{-1},\ldots, \gamma_{Pk}^{-1})$
\State $S_k \leftarrow X^T \Omega_k X + \lambda^2\Gamma_k^{-1}$
\State $\beta_k \leftarrow S_k^{-1} d_k$ Solve the system by conjugate gradient
\EndFor
\Until $\beta$ converges.
\end{algorithmic}
\caption{Batch EM with lasso prior for multinomial logistic regression.}
\label{alg:multinomial AUG CG}
\end{algorithm}

\end{document}